\documentclass[journal=jacsat,manuscript=article]{achemso}

\usepackage{chemformula} 
\usepackage[T1]{fontenc} 
\usepackage{amssymb}
\usepackage{float}
\usepackage[section]{placeins}

\DeclareUnicodeCharacter{2212}{\ensuremath{-}}

\title{Decoupled anisotropic Charge-Phonon Transport Enables Exceptional n-Type Thermoelectric Performance in CuBiSCl$_2$}

\author{Yu Wu}
\email{wuyu9573@qq.com}
\affiliation{Micro- and Nano-scale Thermal Measurement and Thermal Management Laboratory, School of Energy and Mechanical Engineering, Nanjing Normal University, Jiangsu, Nanjing 210023, China}

\author{Ying Chen}
\affiliation{Eastern Institute of Technology, Zhejiang, Ningbo 315200, China}

\author{Shuming Zeng}
\affiliation{College of Physics Science and Technology, Yangzhou University, Jiangsu 225009, China}

\author{Liujiang Zhou}
\affiliation{School of Physics and State Key Laboratory of Electronic Thin Films and Integrated Devices, University of Electronic Science and Technology, Sichuan, Chengdu 610054, China}

\author{Chenhan Liu}
\email{chenhanliu@njnu.edu.cn}
\affiliation{Micro- and Nano-scale Thermal Measurement and Thermal Management Laboratory, School of Energy and Mechanical Engineering, Nanjing Normal University, Jiangsu, Nanjing 210023, China}

\begin{document}

\begin{abstract}

First-principles calculations demonstrate an exceptional decoupling of charge and thermal transport along the \textit{a}-axis in CuBiSCl$_2$. The material achieves superior electron mobility (138 cm$^2$/V$\cdot$s at 300 K) through delocalized Bi-6\textit{p}/S-3\textit{p} networks while maintaining ultralow lattice thermal conductivity (0.40 W/mK at 300 K) via Cu-dominated anharmonic phonon scattering - both optimized along the same crystallographic direction. This simultaneous optimization originates from the anisotropic bonding hierarchy where [BiSCl$_2$]$_n$ ribbons enable efficient charge transport along \textit{a}-axis, while the soft vibrational modes associated with Cu atoms strongly scatter heat-carrying phonons. The resulting high power factor (1.71 mW/mK$^2$ at 700 K) and peak \textit{ZT} of 1.57 establish CuBiSCl$_2$ as a model system that realizes the long-sought "phonon glass-electron crystal" paradigm through crystallographically engineered transport channels.

\end{abstract}

\flushbottom
\maketitle

\thispagestyle{empty}

\section*{Introduction}

Thermoelectric materials enable direct energy conversion between heat and electricity through the Seebeck and Peltier effects, offering promising solutions for waste heat recovery and solid-state refrigeration\cite{Biswas2012,Tan2016,Zhao2014,Wu2024c}. The conversion efficiency is governed by the dimensionless figure of merit $ZT = S^2\sigma T/\kappa_{\text{tot}}$, where $S$, $\sigma$, $T$, and $\kappa_{\text{tot}}$ represent the Seebeck coefficient, electrical conductivity, absolute temperature, and total thermal conductivity (electronic $\kappa_e$ + lattice $\kappa_L$), respectively. Optimizing $ZT$ requires simultaneously enhancing the power factor ($PF = S^2\sigma$) while minimizing $\kappa_{\text{tot}}$, which presents a fundamental challenge due to the strong coupling between these parameters\cite{Biswas2012,Snyder2008,pei2011}. This challenge has motivated the development of materials exhibiting the ``phonon glass-electron crystal'' (PGEC) behavior\cite{Beekman2015}, where glass-like thermal transport coexists with crystal-like electronic properties. Recent advances have demonstrated that anisotropic crystals can achieve this PGEC behavior directionally, with specific crystallographic axes showing simultaneous phonon localization and efficient charge transport\cite{Zhan2024,Zhan2022,Su2022,Wang2023b}.

The fundamental origin of anisotropic transport decoupling lies in the distinct crystallographic dependence of electronic and vibrational properties. In PbSnS$_2$, the out-of-plane transport superiority originates from: (1) interlayer Sn-5\textit{s}/S-3\textit{p} orbital hybridization enabling high electron mobility ($\mu\approx60$ cm$^2$/V$\cdot$s) along the c-axis, and (2) twisted NaCl-type structure with strong anharmonicity suppressing $\kappa_L$ (<0.5 W/m$\cdot$K) in the same direction\cite{Zhan2024}. Parallel mechanisms are observed in n-type SnSe\cite{Su2022}, where (1) strain-induced bond angle reduction enhances out-of-plane charge transport through improved orbital overlap, while (2) the inherent van der Waals gaps between layers establish two-dimensional phonon confinement perpendicular to the conduction pathway. These systems exemplify how directional selectivity can be engineered through either covalent interlayer coupling (PbSnS$_2$) or controlled structural distortion (SnSe), while both approaches utilize anisotropic lattice dynamics to break the traditional transport coupling.

Mixed-anion chalcogenides have recently emerged as promising thermoelectric materials due to their intrinsically low thermal conductivity and tunable electronic properties \cite{Zhao2010,Li2015b,Luu2016}. The coexistence of stereochemically active lone pairs (Bi$^{3+}$ 6\textit{s}$^2$) and mixed anion coordination (S$^{2-}$/Cl$^-$) in compounds like CuBiSCl$_2$ creates a unique bonding hierarchy that strongly suppresses phonon propagation, with reported $\kappa_L$ values below 1 W/m$\cdot$K at room temperature \cite{Shen2024}. However, the fundamental relationship between anisotropic structural motifs and direction-dependent charge transport remains underexplored, particularly regarding how specific crystallographic axes can simultaneously optimize electronic mobility and phonon localization. Moreover, the potential role of four-phonon scattering in such anisotropic systems has not been quantitatively assessed, despite its recognized importance in materials with ultralow thermal conductivity \cite{Wu2024b,Yuan2024}.

In this work, we employ first-principles calculations to systematically investigate the anisotropic thermoelectric properties of CuBiSCl$_2$. Our analysis reveals that the material achieves exceptional \textit{a}-axis performance through: (1) high electron mobility (138 cm$^2$/V$\cdot$s) enabled by delocalized Bi-6\textit{p}/S-3\textit{p} orbitals at the conduction band minimum, and (2) suppressed lattice thermal conductivity ($\kappa_L$ = 0.40 W/m$\cdot$K at 300 K) resulting from Cu-dominated anharmonic vibrations and four-phonon scattering. The combination of these effects yields a peak \textit{ZT} of 1.57 along the \textit{a}-axis at 700 K, establishing CuBiSCl$_2$ as a promising anisotropic thermoelectric material. Our findings provide fundamental insights into how directional bonding hierarchies can be engineered to achieve decoupled charge-phonon transport in mixed-anion compounds.

\section*{Results and Discussion}

\begin{figure*}[ht!]
\centering
\includegraphics[width=1\linewidth]{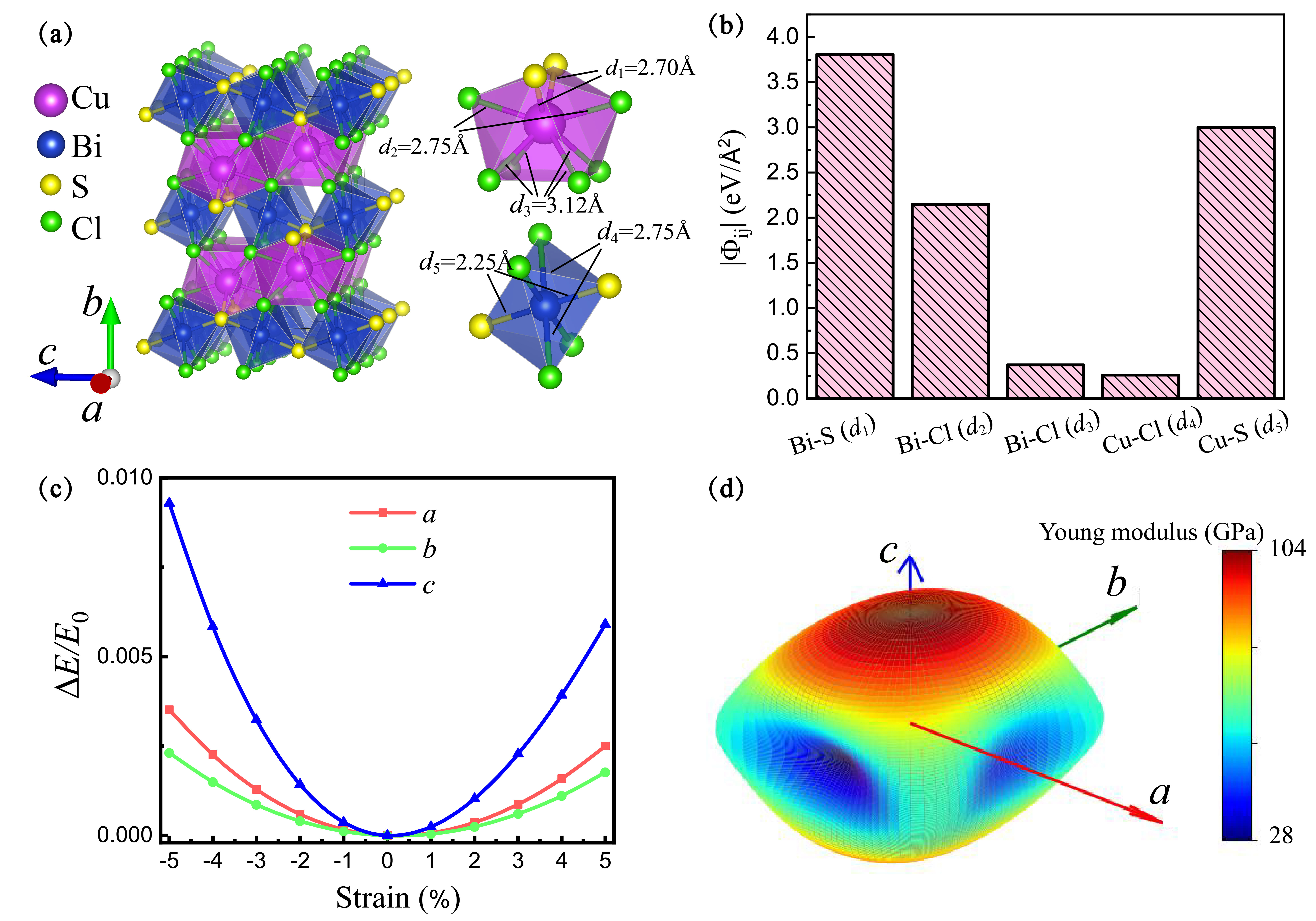}
\caption{(a) The crystal structure of CuBiSCl$_2$ consists of interconnected BiS$_2$Cl$_6$ polyhedra (purple) and CuS$_2$Cl$_4$ octahedra (light blue). Atoms are color-coded: Cu (blue), Bi (purple), S (yellow), and Cl (green). (b) Second-order interatomic force constants for the typical atomic pairs at 300 K. (c) The rate of energy change in the CuBiSCl$_2$ system under strain along different directions. (d) 3D representations of Young's modulus variations.}
\label{Fig1}
\end{figure*}

The crystal structure of CuBiSCl$_2$ (space group \textit{Cmcm}) exhibits pronounced anisotropic bonding through two distinct coordination geometries (Fig.~\ref{Fig1}). Copper ions occupy distorted CuS$_2$Cl$_4$ octahedra, where the axial Cu--S bonds (2.25~\AA) are significantly shorter than the equatorial Cu--Cl distances (2.75~\AA), resulting in a quasi-linear S--Cu--S coordination. This creates a striking contrast between the strong covalent Cu--S interactions (second-order interatomic force constants, IFC = 3.00~eV/\AA$^2$) and weak ionic Cu--Cl contacts (IFC = 0.26~eV/\AA$^2$). Bismuth ions reside in asymmetric BiS$_2$Cl$_6$ polyhedra, displaying three distinct bonding regimes: two short Bi--S bonds (2.70~\AA), two intermediate Bi--Cl bonds (2.75~\AA), and four elongated Bi$\cdots$Cl interactions (3.12~\AA).

The connectivity of these polyhedral units reveals a dimensional hierarchy in bonding strength. Corner-sharing CuS$_2$Cl$_4$ octahedra form [Cu$_2$S$_2$]$_n$ chains along the \textit{c}-axis, while edge-sharing BiS$_2$Cl$_6$ units create [BiSCl$_2$]$_n$ ribbons along the \textit{a}-axis. These strongly bonded frameworks are interconnected along the \textit{b}-axis through weak Cl-mediated linkages (Cu--Cl = 2.75~\AA, Bi$\cdots$Cl = 3.12~\AA), resulting in a ``stiff chain/soft interchain'' architecture. This anisotropic bonding configuration, featuring rigid covalent frameworks separated by weakly bonded chlorine layers, establishes a structural motif that profoundly influences the material's thermal transport characteristics.

The directional mechanical response was characterized through harmonic analysis of strain-energy relationships (Fig.~\ref{Fig1}(c)). Quadratic fitting of the normalized energy variation ($\Delta E/E_0$) versus strain ($\varepsilon$) yields dimensionless coefficients  $\alpha_a$ = 1.2$\times10^{-4}$ (along \textit{a}), $\alpha_b$ = 8.2$\times10^{-5}$ (along \textit{b}) and $\alpha_c$ = 3.0$\times10^{-4}$ (along \textit{c}), revealing a clear stiffness ordering $c > a > b$. This ordering is fully consistent with the 3D Young's modulus results shown in Fig.~\ref{Fig1}(d), where $E_a$ = 40.4~GPa, $E_b$ = 28.0~GPa, and $E_c$ = 104.0~GPa. The exceptional compliance along \textit{b} arises from its unique structural role as the interchain ``soft axis'', where only weak Bi$\cdots$Cl and Cu$\cdots$Cl interactions bridge the rigid [Cu$_2$S$_2$]$_n$ chains (aligned with \textit{c}) and [BiSCl$_2$]$_n$ chains (along \textit{a}).

\begin{figure*}[ht!]
\centering
\includegraphics[width=1\linewidth]{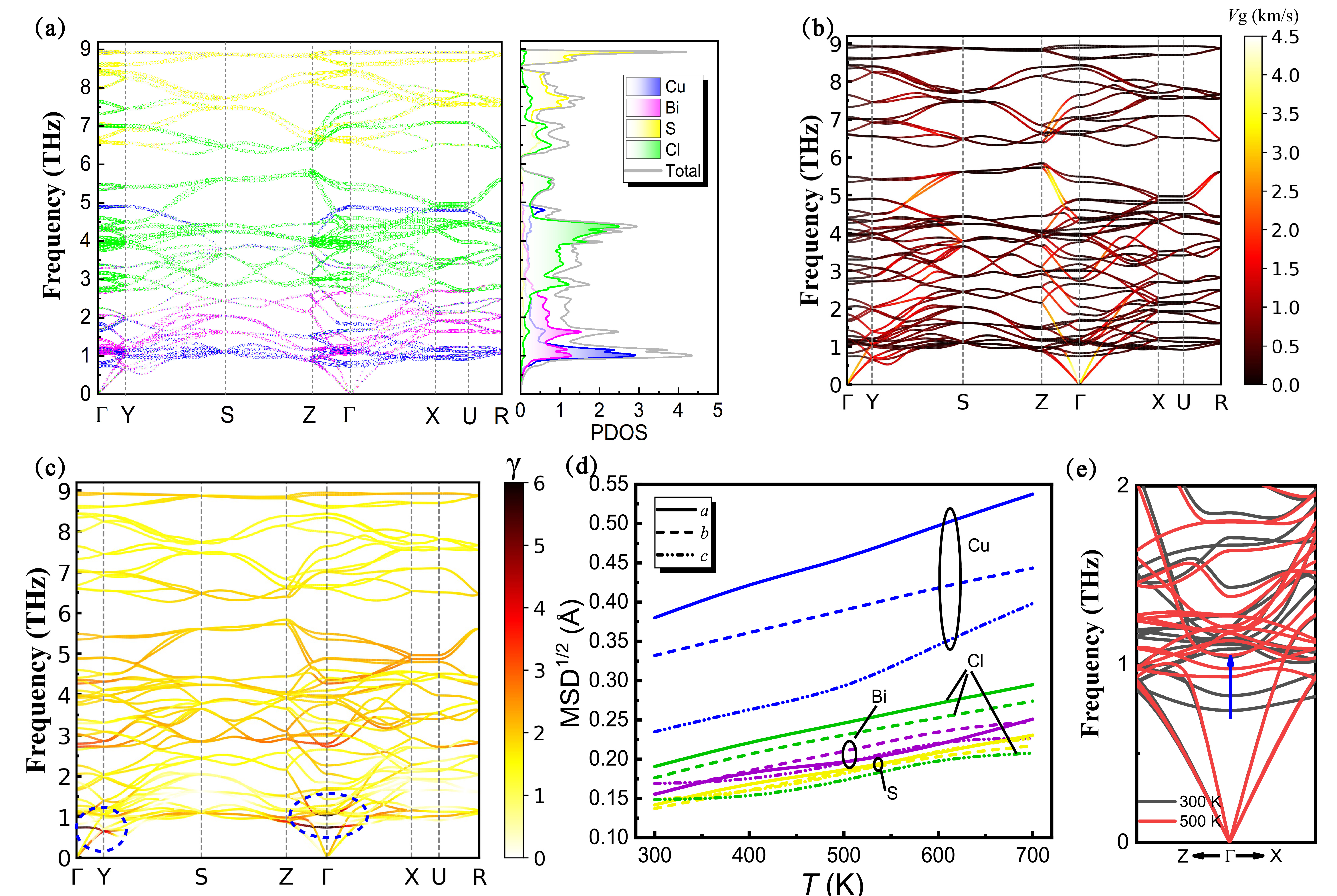}
\caption{(a) The projected phonon dispersion of CuBiSCl$_2$ weighted by their constituted atoms and atom-resolved phonon density of states at 300 K. (b) Phonon dispersion of CuBiSCl$_2$ at 300 K with (b) group velocities and (c) Gr\"uneisen parameters projection. (d) Directional mean-square atomic displacements versus temperature. (e) Phonon dispersion in the low-frequency regime at 300 K and 500 K.}
\label{Fig2}
\end{figure*}

Figure \ref{Fig2}(a) shows the projection of the phonon eigenvectors of each atom onto the phonon dispersion. In reciprocal space, the high-symmetry directions $\Gamma\to\rm{X}$, $\Gamma\to\rm{Y}$, and $\Gamma\to\rm{Z}$ correspond to the real-space crystallographic axes \textit{a}, \textit{b}, and \textit{c}, respectively. The flat phonon dispersion along the $\Gamma$-Y direction, accompanied by significantly reduced group velocities (Fig.~\ref{Fig2}(b)), directly reflects the weak interlayer bonding along the $b$-axis. Due to the strongest stiffness in the $c$ direction, phonons along the $\Gamma$-Z direction have the highest group velocities. The phonon dispersion originating from Cu atomic vibrations exhibits localized modes centered around 1 THz, manifesting as a distinct peak in the phonon density of states (phDOS), with three characteristic signatures: (1) exceptionally low group velocities ($<1$~km/s) persisting throughout the Brillouin zone, (2) strongly enhanced Gr\"uneisen parameters ($\gamma > 5$) near $\Gamma$ indicating giant anharmonicity as shown in Fig.~\ref{Fig2}(c), and (3) anisotropic mean square displacements (MSD) ($\sqrt{\rm{MSD}}_a$ = 0.38 \AA, $\sqrt{\rm{MSD}}_b$ = 0.33 \AA, $\sqrt{\rm{MSD}}_c$ = 0.23 \AA$\;$at 300 K) revealing preferential $ab$-plane vibrations (Fig.~2(d)). These phenomena originate from the quasi-1D S-Cu-S coordination that confines motion along \textit{c}, while weak Cu-Cl interactions enable large-amplitude \textit{ab}-plane displacements. These delocalized \textit{ab}-plane vibrations can strongly couple with and scatter the propagating acoustic phonon modes in the $ab$-plane, significantly suppressing the $ab$-plane lattice thermal conductivity\cite{Zhu2023a}. The observed phonon modes, characterized by their localized nature in frequency space but delocalized behavior in real space, typically exhibit anomalous hardening behavior with increasing temperature as shown in Fig.~\ref{Fig2}(e)\cite{Wu2024a,Thakur2023,Dutta2020}. 

\begin{figure*}[ht!]
\centering
\includegraphics[width=1\linewidth]{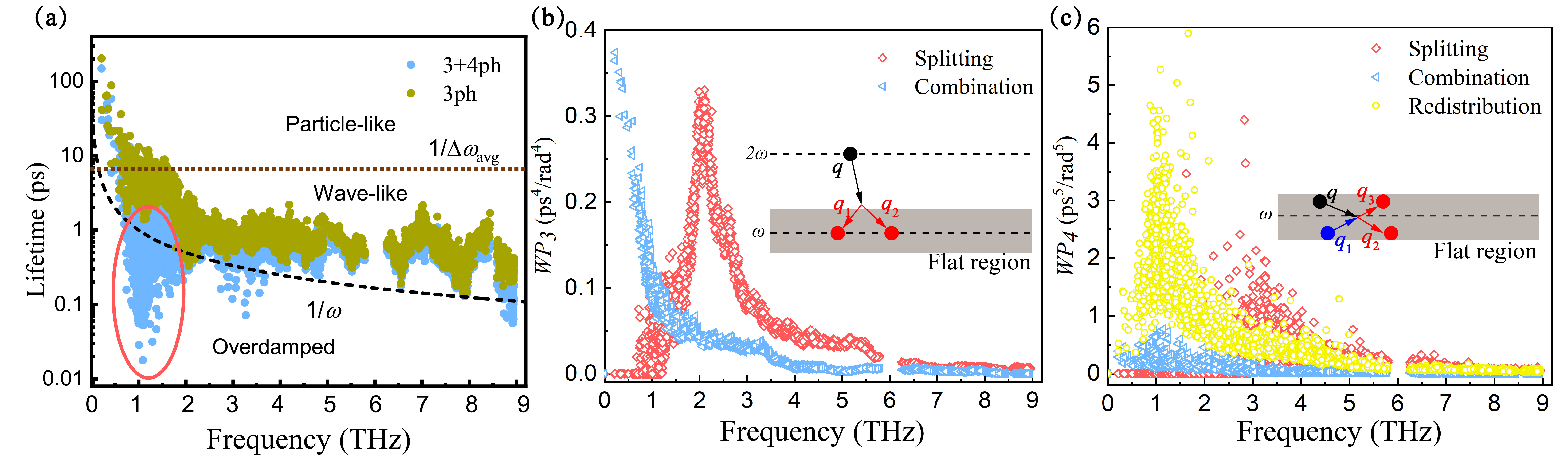}
\caption{(a) Comparison of phonon lifetimes in CuBiSCl$_2$ with 3ph and 3ph+4ph scattering at 300 K. The dashed lines indicate the Ioffe-Regel ($\tau=\omega^{-1}$) and Wigner ($\tau=\Delta\omega_{\rm avg}^{-1}$) limits that separate different heat transport regimes.}
\label{Fig3}
\end{figure*}

The phonon lifetime ($\tau$) analysis of CuBiSCl$_2$ reveals three distinct thermal transport regimes. As shown in Fig.~\ref{Fig3}(a), phonons above the Wigner limit ($\tau > \Delta\omega_{\rm avg}^{-1}$) exhibit particle-like propagation ($\kappa_{\rm p}$), while those between the Wigner and Ioffe-Regel limits ($\omega^{-1} < \tau < \Delta\omega_{\rm avg}^{-1}$) demonstrate wave-like tunneling ($\kappa_{\rm c}$)\cite{DiLucente2023}. Notably, the inclusion of four-phonon scattering induces a sharp reduction in $\tau$ near 1 THz. This frequency corresponds to Cu-dominated vibrations, and the resulting $\tau$ values fall below the Ioffe-Regel limit, placing these modes firmly in the overdamped regime. This behavior indicates that Cu-related phonons undergo extreme anharmonic scattering, transitioning from propagating quasiparticles to localized excitations. To further understand the microscope mechanism of the phonon lifetime reduction in CuBiSCl$_2$ due to four-phonon scattering, we calculated the 3ph ($WP_3$) and 4ph ($WP_4$) scattering phase space, as shown in Fig.~\ref{Fig3}(b, c). The three-phonon emission peak at $2\omega$ in CuBiSCl$_2$ originates from the similar physical mechanism observed in YbFe$_4$Sb$_{12}$\cite{Li2015a}, where flat phonon modes create enhanced scattering channels. For phonons with frequency $2\omega$, they efficiently decay into pairs of Cu-localized $\omega$ modes through the process $q \rightarrow q_1 + q_2$. This is enabled by the combination of strict energy conservation and relaxed momentum selection rules due to the flat modes contributed by Cu vibrations. The $\omega$-peak in four-phonon scattering originates from redistribution processes $q + q_1 \rightarrow q_2 + q_3$\cite{Li2024}. The Cu-derived flat phonon modes relax momentum selection rules, enhancing four-phonon redistribution processes similar to their effect on three-phonon scattering. This leads to anomalously low phonon lifetimes near $\omega$ when four-phonon interactions are included.

\begin{figure*}[ht!]
\centering
\includegraphics[width=1\linewidth]{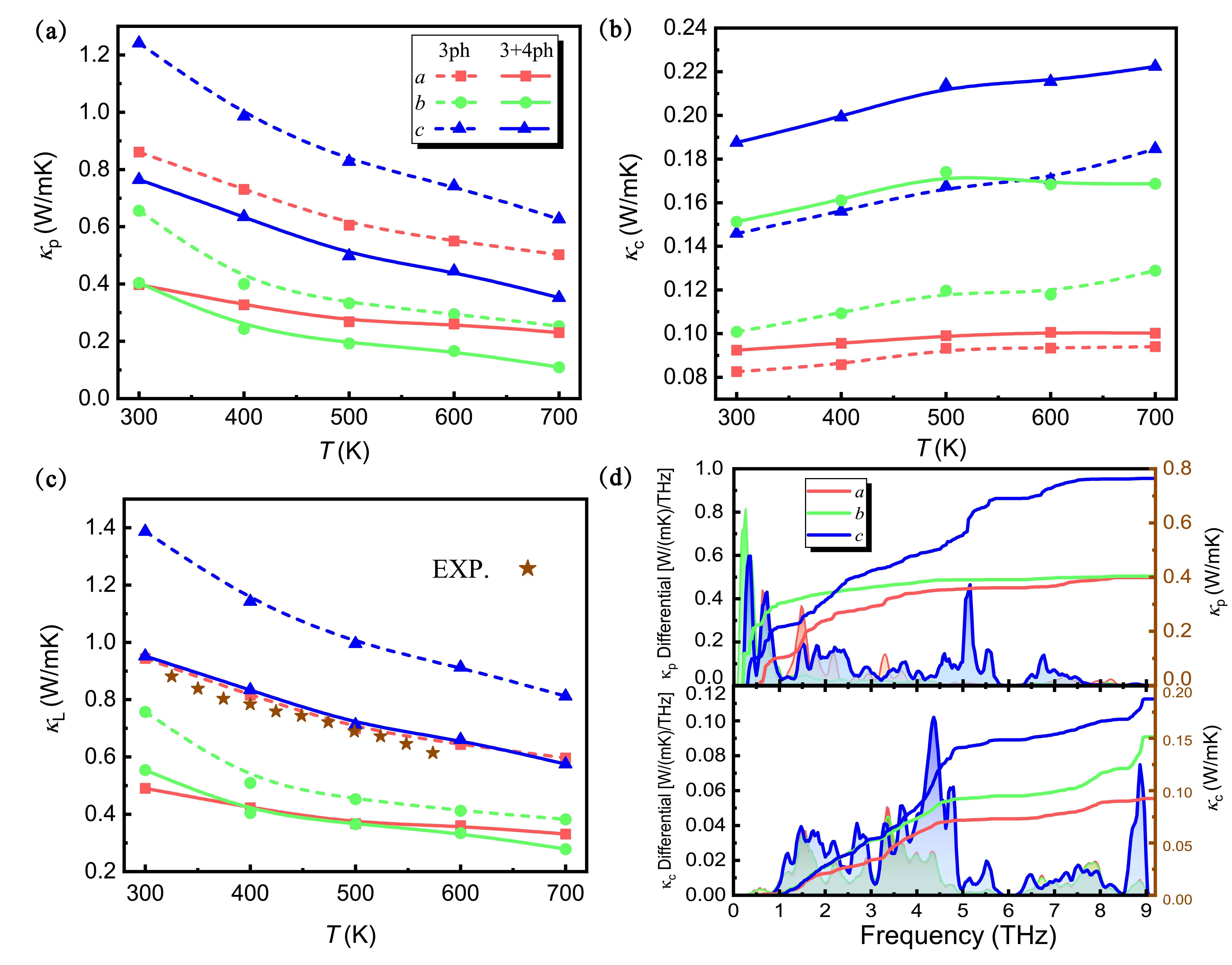}
\caption{Calculated temperature-dependent (a) $\kappa_p$ (b) $\kappa_c$ and (c) $\kappa_L$ considering 3ph and 3+4ph scattering along different directions. (d) The cumulative and differential $\kappa_p$ and $\kappa_c$ as a function of phonon frequencies at 300 K. Experimental data (stars) are from Ref. \cite{Shen2024}.}
\label{Fig4}
\end{figure*}

Figure 4 presents the anisotropic lattice thermal transport behavior in CuBiSCl$_2$, revealing the distinct contributions of particle-like and wave-like phonon transport. Figure \ref{Fig4}(a) demonstrates that four-phonon scattering reduces the particle-like thermal conductivity ($\kappa_p$) across all directions, as exemplified by the $a$-axis decrease from 0.86 W/mK (3ph) to 0.40 W/mK (3+4ph) at 300 K, representing a 53.5\% reduction. The directional dependence of $\kappa_p$ maintains the $c > a > b$ hierarchy, with the $c$-axis $\kappa_p$ (0.76 W/mK) being 1.9 times the $b$-axis value (0.40 W/mK) at 300 K. Figure \ref{Fig4}(b) demonstrates the temperature-dependent enhancement of wave-like thermal conductivity ($\kappa_c$), which reaches 0.22 W/mK (38\% of total $\kappa_L$) along the $c$-axis at 700 K due to increased wave tunneling resulting from four-phonon-induced reduction in phonon lifetimes\cite{Ji2024}. The calculated $\kappa_L$ shows excellent agreement with experimental measurements (Fig.~4(c))\cite{Shen2024}, with the $c$-axis value being significantly larger than those along $a$- and $b$-axes, while the latter two remain comparable in magnitude. Figure 4(d) shows the cumulative and differential $\kappa_p$ and $\kappa_c$ as a function of phonon frequencies at 300 K. The $\kappa_p$ is primarily contributed by low-frequency acoustic phonons with strong dispersion. Along the $c$-axis, optical phonon branches with significant dispersion also make substantial contributions to $\kappa_p$. Interestingly, the frequency range that dominates the $\kappa_c$ corresponds to phonon modes that contribute minimally to $\kappa_p$, and these modes are located in regions of high phDOS. This complementary relationship between $\kappa_p$ and $\kappa_c$ arises because: (1) the strongly dispersive acoustic/optical phonons that enhance $\kappa_p$ have extended propagation characteristics unfavorable for wave tunneling, while (2) the flat optical modes that boost $\kappa_c$ reside in high phDOS regions where multiple scattering channels strongly suppress particle-like thermal transport.

\begin{figure*}[ht!]
\centering
\includegraphics[width=1\linewidth]{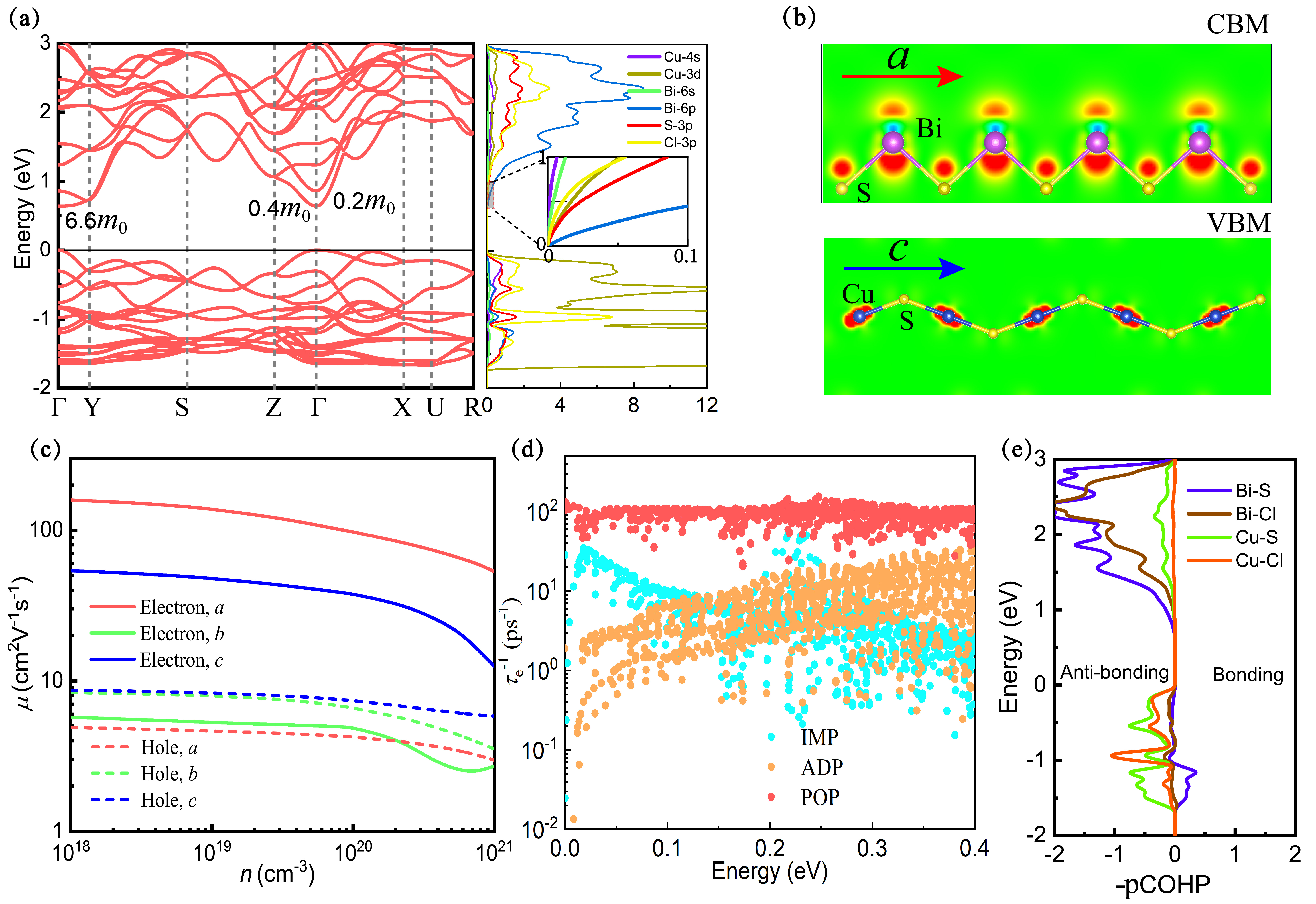}
\caption{(a) The electronic structure and partial electronic density of states (PDOS) for CuBiSCl$_2$. (b) CBM- and VBM-associated partial charge density maps. (c) The variation of electron and hole mobility with concentration along different directions at 300 K. (d) The electron scattering rate near the CBM with $n=10^{19}\;\rm{cm^{-3}}$ and $T=300\;\rm{K}$. (e) The projected crystal orbital Hamilton population (pCOHP).}
\label{Fig5}
\end{figure*}

Figure 5 presents the first-principles analysis of electronic transport properties. The calculated band structure (Fig. 5(a)) reveals a direct bandgap at $\Gamma$ with both conduction band minimum (CBM) and valence band maximum (VBM) located at this high-symmetry point. The electron effective masses along $\Gamma$-X, $\Gamma$-Y, and $\Gamma$-Z directions are 0.2$m_0$, 6.6$m_0$, and 0.4$m_0$ respectively ($m_0$: free electron mass), indicating strongly anisotropic transport favoring the $\Gamma$-X direction (real-space $a$-axis). The projected density of states (PDOS) shows the CBM is primarily composed of Bi 6$p$ and S 3$p$ orbitals, while the VBM derives mainly from Cu 3$d$ orbitals with minor Cl 3$p$ contributions\cite{Xia2025,Ming2022}. This orbital character is visualized in the charge density plots (Fig. 5(b)): the CBM exhibits extended charge density along Bi-S chains, facilitating electron transport, whereas the VBM shows localized charge within Cu-centered octahedra. The directional carrier mobilities at 300 K (Fig. 5(c)) quantitatively reflect this anisotropy, with electron mobilities of 138, 5.3, and 47.9 cm$^2$/V$\cdot$s along $a$-, $b$-, and $c$-axes respectively at 10$^{19}$ cm$^{-3}$ carrier concentration. The mobility reduction at higher concentrations arises from enhanced ionized impurity scattering. The exceptional $a$-axis electron mobility correlates with the delocalized Bi 6$p$/S 3$p$ network observed in the CBM charge distribution. The temperature-dependent mobilities at $n=10^{19}\rm{cm^{-3}}$ are shown in Fig.~S1. Figure \ref{Fig5}(d) reveals the contributions of three scattering mechanisms to the total electron scattering rate: ionized impurity (IMP) scattering, acoustic deformation potential (ADP) scattering, and polar optical phonon (POP) scattering, with POP dominating at approximately 100 ps$^{-1}$. The dominant hole scattering mechanism remains POP scattering, as evidenced by the scattering rate analysis presented in Fig.~S2. Figure \ref{Fig5}(e) presents the projected crystal orbital Hamilton population (pCOHP) analysis, showing predominantly antibonding states near the valence band maximum, which significantly enhances the lattice anharmonicity through weakened interatomic interactions.

\begin{figure*}[ht!]
\centering
\includegraphics[width=1\linewidth]{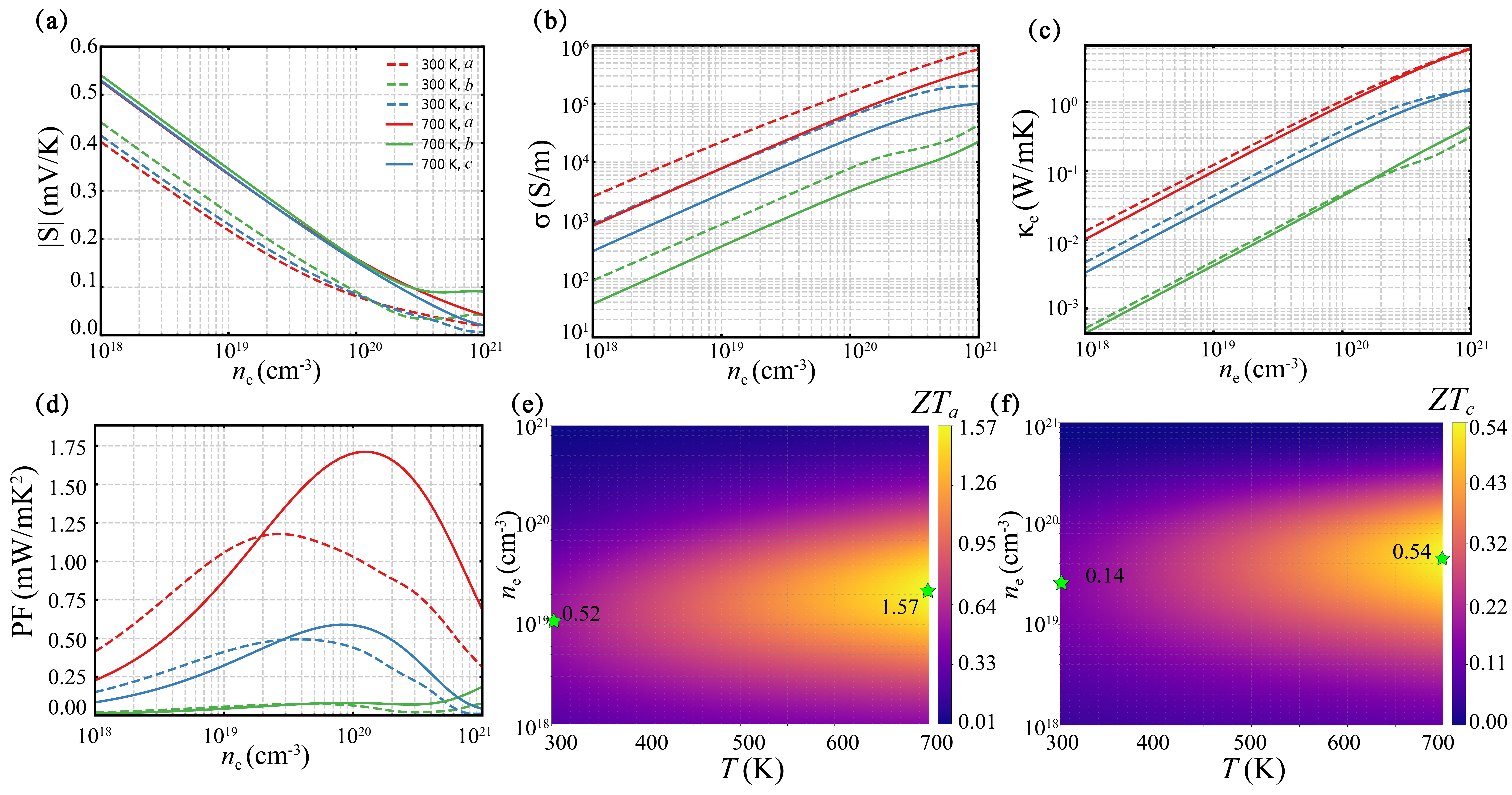}
\caption{The calculated thermoelectric parameters for n-type CuBiSCl$_2$ at 300 and 700 K, respectively. (a) The Seebeck coefficient $|S|$. (b) The electrical conductivity $\sigma$. (c) The electronic thermal conductivity $\kappa_e$. (d) The power factor PF. Concentration- and temperature-dependent $ZT$ along (e) $a$-axis and (f) $c$-axis.}
\label{Fig6}
\end{figure*}

First-principles calculations reveal a remarkable decoupling of charge and thermal transport in CuBiSCl$_2$, suggesting promising thermoelectric potential. The Seebeck coefficient ($|S|$) in Fig.~\ref{Fig6}(a) shows moderate anisotropy with $b$-axis superiority due to flat valence bands and heavy effective mass, while $a$- and $c$-axes exhibit comparable values, indicating relatively weak band anisotropy effects on thermopower. In contrast, electrical conductivity (Fig. 6(b)) displays significantly stronger directional dependence, maintaining the anisotropic hierarchy (\textit{a} > \textit{c} > \textit{b}) consistent with carrier mobility trends, with $\sigma_a$ reaching $2.4\times10^{4}$ S/m at $n=10^{19}\;\rm{cm^{-3}}$ and $T=300\;\rm{K}$. The electronic thermal conductivity ($\kappa_e$) in Fig.~\ref{Fig6}(c) follows a similar directional dependence as the electrical conductivity. This transport behavior yields exceptional \textit{a}-axis power factors (PF) in Fig. 6(d) of 1.18 mW/mK$^2$ (300K) and 1.71 mW/mK$^2$ (700K). The $ZT$ maps in Fig.~\ref{Fig6}(e, f) demonstrate \textit{a}-axis dominance (0.52 at 300K, 1.57 at 700K) resulting from the unique combination of enhanced electronic transport and suppressed phonon conduction along the \textit{a}-axis, establishing optimal thermoelectric performance through this transport decoupling mechanism. The thermoelectric properties of p-type CuBiSCl$_2$ are shown in Fig.~S3.

\section*{Conclusions}

This computational study demonstrates that CuBiSCl$_2$ achieves exceptional thermoelectric performance along the \textit{a}-axis through intrinsic anisotropy in both electronic and thermal transport. The directional hierarchy of chemical bonding creates complementary transport channels, where the \textit{a}-axis simultaneously benefits from high electron mobility (138 cm$^2$/V$\cdot$s) due to delocalized Bi-6$p$/S-3$p$ orbitals and low lattice thermal conductivity (0.40 W/mK at 300 K) through strongly anharmonic Cu vibrations and four-phonon scattering. The resulting decoupling yields a maximum \textit{ZT} of 1.57 at 700 K. Moreover, the dense phonon dispersion and ultralow phonon lifetime lead to non-negligible wave-like thermal transport, with \textit{c}-axis $\kappa_c$ contributing 38\% to the total lattice thermal conductivity at 700 K. These findings establish that the strategic alignment of favorable electronic bands with soft vibrational modes along specific crystallographic directions provides an effective pathway for designing high-performance thermoelectric materials without extrinsic nanostructuring.

\section*{Numerical methods}

The calculations are performed using the Vienna Ab Initio Simulation Package (VASP) based on density functional theory (DFT)\cite{Kresse1996}, employing the projector augmented wave (PAW) method with the PBEsol exchange--correlation functional (a revised Perdew-Burke-Ernzerhof generalized gradient approximation for solids)\cite{Perdew2008}. A plane-wave cutoff energy of 500 eV is used. Atomic positions are optimized with an energy convergence criterion of $10^{-5}$ eV between consecutive steps and a maximum Hellmann-Feynman force tolerance of $10^{-3}$ eV/\r{A}. The conventional cell containing 20 atoms was optimized using a $\Gamma$-centered $\mathbf{q}$--mesh of $12 \times 4 \times 6$ in the Brillouin zone.

For the ab initio molecular dynamics (AIMD) simulations, the 4$\times$2$\times$2 supercell (320 atoms) was constructed by expanding the primitive unit cell (20 atoms) with $\Gamma$-centered 1$\times$1$\times$1 $\mathbf{k}$-mesh. The simulations run for 30 ps with a 1 fs timestep. Interatomic force constants (IFCs) are extracted from AIMD trajectories using the temperature-dependent effective potential (TDEP) method\cite{Hellman2013}. Cutoff radii for third-order and fourth-order IFCs are set to 6 \AA$\;$and 4 \AA, respectively. The lattice thermal conductivity ($\kappa_L$) and related parameters, including phonon relaxation times, are computed using the ShengBTE software\cite{shengbte2014,Han2022}, which implements an iterative solution scheme. A $\mathbf{q}$-mesh of 8$\times$8$\times$8 is adopted in the first irreducible Brillouin zone, with a Gaussian smearing width of 0.2. The four-phonon (4ph) scattering rate is efficiently calculated using a maximum likelihood estimation method\cite{Guo2024} with a sample size of $4\times10^{5}$.The electronic transport properties, including electrical conductivity ($\sigma$), Seebeck coefficient ($S$), and electronic thermal conductivity ($\kappa_e$), are computed using the AMSET code~\cite{Ganose2021}. The considered scattering mechanisms include ionized impurity (IMP), acoustic deformation potential (ADP), and polar optical phonon (POP) scattering. The hybrid HSE06 functional\cite{Heyd2003}, which incorporates spin-orbit coupling (SOC), is employed to obtain an accurate band gap of 1.3 eV, with a screening parameter set to 0.25 \AA$^{-1}$.
.

\section*{Conflicts of interest}

There are no conflicts to declare.

\section*{Acknowledgements}

This work is supported by the Natural Science Foundation of China (12304038, 52206092), the Startup funds of Outstanding Talents of UESTC (A1098531023601205), National Youth Talents Plan of China (G05QNQR049), the Big Data Computing Center of Southeast University.



\begin{mcitethebibliography}{36}
\providecommand*\natexlab[1]{#1}
\providecommand*\mciteSetBstSublistMode[1]{}
\providecommand*\mciteSetBstMaxWidthForm[2]{}
\providecommand*\mciteBstWouldAddEndPuncttrue
  {\def\EndOfBibitem{\unskip.}}
\providecommand*\mciteBstWouldAddEndPunctfalse
  {\let\EndOfBibitem\relax}
\providecommand*\mciteSetBstMidEndSepPunct[3]{}
\providecommand*\mciteSetBstSublistLabelBeginEnd[3]{}
\providecommand*\EndOfBibitem{}
\mciteSetBstSublistMode{f}
\mciteSetBstMaxWidthForm{subitem}{(\alph{mcitesubitemcount})}
\mciteSetBstSublistLabelBeginEnd
  {\mcitemaxwidthsubitemform\space}
  {\relax}
  {\relax}

\bibitem[Biswas \latin{et~al.}(2012)Biswas, He, Blum, Wu, Hogan, Seidman,
  Dravid, and Kanatzidis]{Biswas2012}
Biswas,~K.; He,~J.; Blum,~I.~D.; Wu,~C.-I.; Hogan,~T.~P.; Seidman,~D.~N.;
  Dravid,~V.~P.; Kanatzidis,~M.~G. High-performance bulk thermoelectrics with
  all-scale hierarchical architectures. \emph{Nature} \textbf{2012},
  \emph{489}, 414--418\relax
\mciteBstWouldAddEndPuncttrue
\mciteSetBstMidEndSepPunct{\mcitedefaultmidpunct}
{\mcitedefaultendpunct}{\mcitedefaultseppunct}\relax
\EndOfBibitem
\bibitem[Tan \latin{et~al.}(2016)Tan, Zhao, and Kanatzidis]{Tan2016}
Tan,~G.; Zhao,~L.-D.; Kanatzidis,~M.~G. Rationally designing high-performance
  bulk thermoelectric materials. \emph{Chem. Rev.} \textbf{2016}, \emph{116},
  12123--12149\relax
\mciteBstWouldAddEndPuncttrue
\mciteSetBstMidEndSepPunct{\mcitedefaultmidpunct}
{\mcitedefaultendpunct}{\mcitedefaultseppunct}\relax
\EndOfBibitem
\bibitem[Zhao \latin{et~al.}(2014)Zhao, Lo, Zhang, Sun, Tan, Uher, Wolverton,
  Dravid, and Kanatzidis]{Zhao2014}
Zhao,~L.-D.; Lo,~S.-H.; Zhang,~Y.; Sun,~H.; Tan,~G.; Uher,~C.; Wolverton,~C.;
  Dravid,~V.~P.; Kanatzidis,~M.~G. {Ultralow thermal conductivity and high
  thermoelectric figure of merit in SnSe crystals.} \emph{Nature}
  \textbf{2014}, \emph{508}, 373--7\relax
\mciteBstWouldAddEndPuncttrue
\mciteSetBstMidEndSepPunct{\mcitedefaultmidpunct}
{\mcitedefaultendpunct}{\mcitedefaultseppunct}\relax
\EndOfBibitem
\bibitem[Wu \latin{et~al.}(2024)Wu, Ji, Ding, and Zhou]{Wu2024c}
Wu,~Y.; Ji,~L.; Ding,~Y.; Zhou,~L. High throughput screening of semiconductors
  with low lattice thermal transport induced by long-range interactions.
  \emph{Mater. Horiz.} \textbf{2024}, \emph{11}, 3651--3661\relax
\mciteBstWouldAddEndPuncttrue
\mciteSetBstMidEndSepPunct{\mcitedefaultmidpunct}
{\mcitedefaultendpunct}{\mcitedefaultseppunct}\relax
\EndOfBibitem
\bibitem[Snyder and Toberer(2008)Snyder, and Toberer]{Snyder2008}
Snyder,~G.~J.; Toberer,~E.~S. Complex thermoelectric materials. \emph{Nat.
  Mater.} \textbf{2008}, \emph{7}, 105--114\relax
\mciteBstWouldAddEndPuncttrue
\mciteSetBstMidEndSepPunct{\mcitedefaultmidpunct}
{\mcitedefaultendpunct}{\mcitedefaultseppunct}\relax
\EndOfBibitem
\bibitem[Pei \latin{et~al.}({2011})Pei, Shi, LaLonde, Wang, Chen, and
  Snyder]{pei2011}
Pei,~Y.; Shi,~X.; LaLonde,~A.; Wang,~H.; Chen,~L.; Snyder,~G.~J. {Convergence
  of electronic bands for high performance bulk thermoelectrics}. \emph{Nature}
  \textbf{{2011}}, \emph{{473}}, {66--69}\relax
\mciteBstWouldAddEndPuncttrue
\mciteSetBstMidEndSepPunct{\mcitedefaultmidpunct}
{\mcitedefaultendpunct}{\mcitedefaultseppunct}\relax
\EndOfBibitem
\bibitem[Beekman \latin{et~al.}(2015)Beekman, Morelli, and Nolas]{Beekman2015}
Beekman,~M.; Morelli,~D.~T.; Nolas,~G.~S. Better thermoelectrics through
  glass-like crystals. \emph{Nat. Mater.} \textbf{2015}, \emph{14},
  1182--1185\relax
\mciteBstWouldAddEndPuncttrue
\mciteSetBstMidEndSepPunct{\mcitedefaultmidpunct}
{\mcitedefaultendpunct}{\mcitedefaultseppunct}\relax
\EndOfBibitem
\bibitem[Zhan \latin{et~al.}(2024)Zhan, Bai, Qiu, Zheng, Wang, Zhu, Tan, and
  Zhao]{Zhan2024}
Zhan,~S.; Bai,~S.; Qiu,~Y.; Zheng,~L.; Wang,~S.; Zhu,~Y.; Tan,~Q.; Zhao,~L.
  Insight into carrier and phonon transports of PbSnS2 crystals. \emph{Adv.
  Mater.} \textbf{2024}, \emph{36}\relax
\mciteBstWouldAddEndPuncttrue
\mciteSetBstMidEndSepPunct{\mcitedefaultmidpunct}
{\mcitedefaultendpunct}{\mcitedefaultseppunct}\relax
\EndOfBibitem
\bibitem[Zhan \latin{et~al.}(2022)Zhan, Hong, Qin, Zhu, Feng, Su, Shi, Liang,
  Zhang, Gao, Ge, Zheng, Wang, and Zhao]{Zhan2022}
Zhan,~S.; Hong,~T.; Qin,~B.; Zhu,~Y.; Feng,~X.; Su,~L.; Shi,~H.; Liang,~H.;
  Zhang,~Q.; Gao,~X.; Ge,~Z.-H.; Zheng,~L.; Wang,~D.; Zhao,~L.-D. Realizing
  high-ranged thermoelectric performance in PbSnS2 crystals. \emph{Nat.
  Commun.} \textbf{2022}, \emph{13}\relax
\mciteBstWouldAddEndPuncttrue
\mciteSetBstMidEndSepPunct{\mcitedefaultmidpunct}
{\mcitedefaultendpunct}{\mcitedefaultseppunct}\relax
\EndOfBibitem
\bibitem[Su \latin{et~al.}(2022)Su, Wang, Wang, Qin, Wang, Qin, Jin, Chang, and
  Zhao]{Su2022}
Su,~L.; Wang,~D.; Wang,~S.; Qin,~B.; Wang,~Y.; Qin,~Y.; Jin,~Y.; Chang,~C.;
  Zhao,~L.-D. High thermoelectric performance realized through manipulating
  layered phonon-electron decoupling. \emph{Science} \textbf{2022}, \emph{375},
  1385--1389\relax
\mciteBstWouldAddEndPuncttrue
\mciteSetBstMidEndSepPunct{\mcitedefaultmidpunct}
{\mcitedefaultendpunct}{\mcitedefaultseppunct}\relax
\EndOfBibitem
\bibitem[Wang \latin{et~al.}(2023)Wang, Duan, Zhang, Ma, Zhu, Qian, Yang, Liu,
  and Yang]{Wang2023b}
Wang,~T.; Duan,~X.; Zhang,~H.; Ma,~J.; Zhu,~H.; Qian,~X.; Yang,~J.; Liu,~T.;
  Yang,~R. Origins of three‐dimensional charge and two‐dimensional phonon
  transports in Pnma phase PbSnSe2 thermoelectric crystal. \emph{InfoMat}
  \textbf{2023}, \emph{5}\relax
\mciteBstWouldAddEndPuncttrue
\mciteSetBstMidEndSepPunct{\mcitedefaultmidpunct}
{\mcitedefaultendpunct}{\mcitedefaultseppunct}\relax
\EndOfBibitem
\bibitem[Zhao \latin{et~al.}(2010)Zhao, Berardan, Pei, Byl, Pinsard-Gaudart,
  and Dragoe]{Zhao2010}
Zhao,~L.~D.; Berardan,~D.; Pei,~Y.~L.; Byl,~C.; Pinsard-Gaudart,~L.; Dragoe,~N.
  Bi1−xSrxCuSeO oxyselenides as promising thermoelectric materials.
  \emph{App. Phys. Lett.} \textbf{2010}, \emph{97}\relax
\mciteBstWouldAddEndPuncttrue
\mciteSetBstMidEndSepPunct{\mcitedefaultmidpunct}
{\mcitedefaultendpunct}{\mcitedefaultseppunct}\relax
\EndOfBibitem
\bibitem[Li \latin{et~al.}(2015)Li, Xiao, Fan, Deng, Zhang, Ye, and
  Xie]{Li2015b}
Li,~Z.; Xiao,~C.; Fan,~S.; Deng,~Y.; Zhang,~W.; Ye,~B.; Xie,~Y. Dual vacancies:
  An effective strategy realizing synergistic optimization of thermoelectric
  property in BiCuSeO. \emph{J. Am. Chem. Soc} \textbf{2015}, \emph{137},
  6587--6593\relax
\mciteBstWouldAddEndPuncttrue
\mciteSetBstMidEndSepPunct{\mcitedefaultmidpunct}
{\mcitedefaultendpunct}{\mcitedefaultseppunct}\relax
\EndOfBibitem
\bibitem[Luu and Vaqueiro(2016)Luu, and Vaqueiro]{Luu2016}
Luu,~S.~D.; Vaqueiro,~P. Layered oxychalcogenides: Structural chemistry and
  thermoelectric properties. \emph{J. Materiomics} \textbf{2016}, \emph{2},
  131--140\relax
\mciteBstWouldAddEndPuncttrue
\mciteSetBstMidEndSepPunct{\mcitedefaultmidpunct}
{\mcitedefaultendpunct}{\mcitedefaultseppunct}\relax
\EndOfBibitem
\bibitem[Shen \latin{et~al.}(2024)Shen, Pal, Acharyya, Raveau, Boullay,
  Lebedev, Prestipino, Fujii, Yang, Tsao, Renaud, Lemoine, Candolfi, and
  Guilmeau]{Shen2024}
Shen,~X.; Pal,~K.; Acharyya,~P.; Raveau,~B.; Boullay,~P.; Lebedev,~O.~I.;
  Prestipino,~C.; Fujii,~S.; Yang,~C.-C.; Tsao,~I.-Y.; Renaud,~A.; Lemoine,~P.;
  Candolfi,~C.; Guilmeau,~E. Lone pair induced 1D character and weak
  cation–anion interactions: Two ingredients for low thermal conductivity in
  mixed-anion metal chalcohalide CuBiSCl2. \emph{J. Am. Chem. Soc}
  \textbf{2024}, \emph{146}, 29072--29083\relax
\mciteBstWouldAddEndPuncttrue
\mciteSetBstMidEndSepPunct{\mcitedefaultmidpunct}
{\mcitedefaultendpunct}{\mcitedefaultseppunct}\relax
\EndOfBibitem
\bibitem[Wu \latin{et~al.}(2024)Wu, Dai, Ji, Ding, Zeng, Yang, and
  Zhou]{Wu2024b}
Wu,~Y.; Dai,~S.; Ji,~L.; Ding,~Y.; Zeng,~S.; Yang,~J.; Zhou,~L. Impact of the
  acoustic phonon bandwidth on three-phonon and four-phonon scattering in
  half-Heusler materials. \emph{Phys. Rev. B} \textbf{2024}, \emph{110}\relax
\mciteBstWouldAddEndPuncttrue
\mciteSetBstMidEndSepPunct{\mcitedefaultmidpunct}
{\mcitedefaultendpunct}{\mcitedefaultseppunct}\relax
\EndOfBibitem
\bibitem[Yuan \latin{et~al.}(2024)Yuan, Zhao, Sun, Ni, and Dai]{Yuan2024}
Yuan,~X.; Zhao,~Y.; Sun,~Y.; Ni,~J.; Dai,~Z. Influence of quartic anharmonicity
  on lattice dynamics and thermal transport properties of 16 antifluorite
  structures. \emph{Phys. Rev. B} \textbf{2024}, \emph{110}, 014304\relax
\mciteBstWouldAddEndPuncttrue
\mciteSetBstMidEndSepPunct{\mcitedefaultmidpunct}
{\mcitedefaultendpunct}{\mcitedefaultseppunct}\relax
\EndOfBibitem
\bibitem[Zhu \latin{et~al.}(2023)Zhu, Xie, Ti, Li, Guo, Zhang, Liu, Tao, Liu,
  Zhang, and Sui]{Zhu2023a}
Zhu,~J.; Xie,~L.; Ti,~Z.; Li,~J.; Guo,~M.; Zhang,~X.; Liu,~P.-F.; Tao,~L.;
  Liu,~Z.; Zhang,~Y.; Sui,~J. Computational understanding and prediction of
  8-electron half-Heusler compounds with unusual suppressed phonon conduction.
  \emph{Appl. Phys. Rev.} \textbf{2023}, \emph{10}\relax
\mciteBstWouldAddEndPuncttrue
\mciteSetBstMidEndSepPunct{\mcitedefaultmidpunct}
{\mcitedefaultendpunct}{\mcitedefaultseppunct}\relax
\EndOfBibitem
\bibitem[Wu \latin{et~al.}(2024)Wu, Huang, Ji, Ji, Ding, and Zhou]{Wu2024a}
Wu,~Y.; Huang,~A.; Ji,~L.; Ji,~J.; Ding,~Y.; Zhou,~L. Origin of intrinsically
  low lattice thermal conductivity in solids. \emph{J. Phys. Chem. Lett.}
  \textbf{2024}, 11525--11537\relax
\mciteBstWouldAddEndPuncttrue
\mciteSetBstMidEndSepPunct{\mcitedefaultmidpunct}
{\mcitedefaultendpunct}{\mcitedefaultseppunct}\relax
\EndOfBibitem
\bibitem[Thakur and Giri(2023)Thakur, and Giri]{Thakur2023}
Thakur,~S.; Giri,~A. Origin of ultralow thermal conductivity in metal halide
  perovskites. \emph{ACS Appl. Mater.} \textbf{2023}, \emph{15},
  26755--26765\relax
\mciteBstWouldAddEndPuncttrue
\mciteSetBstMidEndSepPunct{\mcitedefaultmidpunct}
{\mcitedefaultendpunct}{\mcitedefaultseppunct}\relax
\EndOfBibitem
\bibitem[Dutta \latin{et~al.}(2020)Dutta, Samanta, Ghosh, Voneshen, and
  Biswas]{Dutta2020}
Dutta,~M.; Samanta,~M.; Ghosh,~T.; Voneshen,~D.~J.; Biswas,~K. Evidence of
  highly anharmonic soft lattice vibrations in a zintl rattler. \emph{Angew.
  Chem. Int. Ed.} \textbf{2020}, \emph{60}, 4259--4265\relax
\mciteBstWouldAddEndPuncttrue
\mciteSetBstMidEndSepPunct{\mcitedefaultmidpunct}
{\mcitedefaultendpunct}{\mcitedefaultseppunct}\relax
\EndOfBibitem
\bibitem[Di~Lucente \latin{et~al.}(2023)Di~Lucente, Simoncelli, and
  Marzari]{DiLucente2023}
Di~Lucente,~E.; Simoncelli,~M.; Marzari,~N. Crossover from Boltzmann to Wigner
  thermal transport in thermoelectric skutterudites. \emph{Phys. Rev. Res.}
  \textbf{2023}, \emph{5}\relax
\mciteBstWouldAddEndPuncttrue
\mciteSetBstMidEndSepPunct{\mcitedefaultmidpunct}
{\mcitedefaultendpunct}{\mcitedefaultseppunct}\relax
\EndOfBibitem
\bibitem[Li and Mingo(2015)Li, and Mingo]{Li2015a}
Li,~W.; Mingo,~N. Ultralow lattice thermal conductivity of the fully filled
  skutterudite YbFe4Sb12 due to the flat avoided-crossing filler modes.
  \emph{Phys. Rev. B} \textbf{2015}, \emph{91}, 144304\relax
\mciteBstWouldAddEndPuncttrue
\mciteSetBstMidEndSepPunct{\mcitedefaultmidpunct}
{\mcitedefaultendpunct}{\mcitedefaultseppunct}\relax
\EndOfBibitem
\bibitem[Li \latin{et~al.}(2024)Li, Chen, Lu, Fukui, Yu, Li, Zhao, Wang, Wang,
  and Hong]{Li2024}
Li,~Y.; Chen,~J.; Lu,~C.; Fukui,~H.; Yu,~X.; Li,~C.; Zhao,~J.; Wang,~X.;
  Wang,~W.; Hong,~J. Multiphonon interaction and thermal conductivity in
  half-Heusler LuNiBi. \emph{Phys. Rev. B} \textbf{2024}, \emph{109},
  174302\relax
\mciteBstWouldAddEndPuncttrue
\mciteSetBstMidEndSepPunct{\mcitedefaultmidpunct}
{\mcitedefaultendpunct}{\mcitedefaultseppunct}\relax
\EndOfBibitem
\bibitem[Ji \latin{et~al.}(2024)Ji, Huang, Huo, Ding, Zeng, Wu, and
  Zhou]{Ji2024}
Ji,~L.; Huang,~A.; Huo,~Y.; Ding,~Y.-m.; Zeng,~S.; Wu,~Y.; Zhou,~L. Influence
  of four-phonon scattering and wavelike phonon tunneling effects on the
  thermal transport properties of TlBiSe2. \emph{Phys. Rev. B} \textbf{2024},
  \emph{109}, 214307\relax
\mciteBstWouldAddEndPuncttrue
\mciteSetBstMidEndSepPunct{\mcitedefaultmidpunct}
{\mcitedefaultendpunct}{\mcitedefaultseppunct}\relax
\EndOfBibitem
\bibitem[Xia \latin{et~al.}(2025)Xia, Li, Xu, and Niu]{Xia2025}
Xia,~M.; Li,~Y.; Xu,~Y.; Niu,~G. High‐density post‐perovskite for
  ultra‐sensitive hard X‐ray detection. \emph{Angew. Chem. Int. Ed}
  \textbf{2025}, \relax
\mciteBstWouldAddEndPunctfalse
\mciteSetBstMidEndSepPunct{\mcitedefaultmidpunct}
{}{\mcitedefaultseppunct}\relax
\EndOfBibitem
\bibitem[Ming \latin{et~al.}(2022)Ming, Chen, Zhang, Gong, Wu, Jiang, Ye, Xu,
  Yang, Wang, Cao, Yang, Zhang, Zhang, Shi, and Sun]{Ming2022}
Ming,~C. \latin{et~al.}  Mixed chalcogenide‐halides for stable, lead‐free
  and defect‐tolerant photovoltaics: computational screening and experimental
  validation of CuBiSCl2 with ideal band gap. \emph{Adv. Funct. Mater.}
  \textbf{2022}, \emph{32}, 2112682\relax
\mciteBstWouldAddEndPuncttrue
\mciteSetBstMidEndSepPunct{\mcitedefaultmidpunct}
{\mcitedefaultendpunct}{\mcitedefaultseppunct}\relax
\EndOfBibitem
\bibitem[Kresse and Furthmuller({1996})Kresse, and Furthmuller]{Kresse1996}
Kresse,~G.; Furthmuller,~J. {Efficient iterative schemes for ab initio
  total-energy calculations using a plane-wave basis set}. \emph{{Phys. Rev.
  B}} \textbf{{1996}}, \emph{{54}}, {11169--11186}\relax
\mciteBstWouldAddEndPuncttrue
\mciteSetBstMidEndSepPunct{\mcitedefaultmidpunct}
{\mcitedefaultendpunct}{\mcitedefaultseppunct}\relax
\EndOfBibitem
\bibitem[Perdew \latin{et~al.}(2008)Perdew, Ruzsinszky, Csonka, Vydrov,
  Scuseria, Constantin, Zhou, and Burke]{Perdew2008}
Perdew,~J.~P.; Ruzsinszky,~A.; Csonka,~G.~I.; Vydrov,~O.~A.; Scuseria,~G.~E.;
  Constantin,~L.~A.; Zhou,~X.; Burke,~K. Restoring the density-gradient
  expansion for exchange in solids and surfaces. \emph{Phys. Rev. Lett.}
  \textbf{2008}, \emph{100}, 136406\relax
\mciteBstWouldAddEndPuncttrue
\mciteSetBstMidEndSepPunct{\mcitedefaultmidpunct}
{\mcitedefaultendpunct}{\mcitedefaultseppunct}\relax
\EndOfBibitem
\bibitem[Hellman and Abrikosov(2013)Hellman, and Abrikosov]{Hellman2013}
Hellman,~O.; Abrikosov,~I.~A. Temperature-dependent effective third-order
  interatomic force constants from first principles. \emph{Phys. Rev. B}
  \textbf{2013}, \emph{88}, 144301\relax
\mciteBstWouldAddEndPuncttrue
\mciteSetBstMidEndSepPunct{\mcitedefaultmidpunct}
{\mcitedefaultendpunct}{\mcitedefaultseppunct}\relax
\EndOfBibitem
\bibitem[Li \latin{et~al.}({2014})Li, Carrete, Katcho, and Mingo]{shengbte2014}
Li,~W.; Carrete,~J.; Katcho,~N.~A.; Mingo,~N. {ShengBTE: A solver of the
  Boltzmann transport equation for phonons}. \emph{Comput. Phys. Commun.}
  \textbf{{2014}}, \emph{{185}}, {1747--1758}\relax
\mciteBstWouldAddEndPuncttrue
\mciteSetBstMidEndSepPunct{\mcitedefaultmidpunct}
{\mcitedefaultendpunct}{\mcitedefaultseppunct}\relax
\EndOfBibitem
\bibitem[Han \latin{et~al.}(2022)Han, Yang, Li, Feng, and Ruan]{Han2022}
Han,~Z.; Yang,~X.; Li,~W.; Feng,~T.; Ruan,~X. FourPhonon: An extension module
  to ShengBTE for computing four-phonon scattering rates and thermal
  conductivity. \emph{Comput. Phys. Commun.} \textbf{2022}, \emph{270},
  108179\relax
\mciteBstWouldAddEndPuncttrue
\mciteSetBstMidEndSepPunct{\mcitedefaultmidpunct}
{\mcitedefaultendpunct}{\mcitedefaultseppunct}\relax
\EndOfBibitem
\bibitem[Guo \latin{et~al.}(2024)Guo, Han, Feng, Lin, and Ruan]{Guo2024}
Guo,~Z.; Han,~Z.; Feng,~D.; Lin,~G.; Ruan,~X. Sampling-accelerated prediction
  of phonon scattering rates for converged thermal conductivity and radiative
  properties. \emph{npj Comput. Mater.} \textbf{2024}, \emph{10}\relax
\mciteBstWouldAddEndPuncttrue
\mciteSetBstMidEndSepPunct{\mcitedefaultmidpunct}
{\mcitedefaultendpunct}{\mcitedefaultseppunct}\relax
\EndOfBibitem
\bibitem[Ganose \latin{et~al.}(2021)Ganose, Park, Faghaninia, Woods-Robinson,
  Persson, and Jain]{Ganose2021}
Ganose,~A.~M.; Park,~J.; Faghaninia,~A.; Woods-Robinson,~R.; Persson,~K.~A.;
  Jain,~A. Efficient calculation of carrier scattering rates from first
  principles. \emph{Nat. Commun.} \textbf{2021}, \emph{12}, 2222\relax
\mciteBstWouldAddEndPuncttrue
\mciteSetBstMidEndSepPunct{\mcitedefaultmidpunct}
{\mcitedefaultendpunct}{\mcitedefaultseppunct}\relax
\EndOfBibitem
\bibitem[Heyd \latin{et~al.}(2003)Heyd, Scuseria, and Ernzerhof]{Heyd2003}
Heyd,~J.; Scuseria,~G.~E.; Ernzerhof,~M. Hybrid functionals based on a screened
  Coulomb potential. \emph{J. Chem. Phys.} \textbf{2003}, \emph{118},
  8207--8215\relax
\mciteBstWouldAddEndPuncttrue
\mciteSetBstMidEndSepPunct{\mcitedefaultmidpunct}
{\mcitedefaultendpunct}{\mcitedefaultseppunct}\relax
\EndOfBibitem
\end{mcitethebibliography}

\providecommand{\latin}[1]{#1}
\makeatletter
\providecommand{\doi}
  {\begingroup\let\do\@makeother\dospecials
  \catcode`\{=1 \catcode`\}=2 \doi@aux}
\providecommand{\doi@aux}[1]{\endgroup\texttt{#1}}
\makeatother
\providecommand*\mcitethebibliography{\thebibliography}
\csname @ifundefined\endcsname{endmcitethebibliography}
  {\let\endmcitethebibliography\endthebibliography}{}

\end{document}